\documentclass[
    amsmath,
    amssymb,
    aps,
    floatfix,
    twocolumn
]{revtex4-2}
\usepackage[utf8]{inputenc}
\usepackage[colorlinks=true,citecolor=blue,allcolors=blue]{hyperref}
\usepackage{amsmath,amsthm,amsfonts,amssymb,amscd}
\usepackage{graphicx}
\usepackage{lastpage}
\usepackage{enumerate}
\usepackage{mathrsfs}
\usepackage{bm}
\usepackage{braket}
\usepackage[version=4]{mhchem}
\usepackage{blindtext}
\usepackage{natbib}
\usepackage{chemformula}
\usepackage{siunitx}
\usepackage{color,soul}

\newcommand{\target}{HNC$_3$}
\newcommand{\anion}{HNC$_3^-$}

\begin{document}

\author{
    Elizabeth Aubin$^{1}$,
    Jean-Christophe Loison$^2$,
    Mehdi Ayouz$^{3,4}$,
    Joshua Forer$^5$,
    Viatcheslav Kokoouline$^1$
}

\homepage{E-mail: slavako@ucf.edu}
\affiliation{
    $^1$Department of Physics, University of Central Florida, Orlando, FL32816, U.S.A. \\
    $^2$Institut des Sciences Mol\'eculaires, Universit\'e de Bordeaux, France \\
    $^3$Universit\'e Paris-Saclay, CNRS, CentraleSup\'elec, Structures Propriétés et Modélisation des solides, Gif-sur-Yvette, France \\
    $^4$Universit\'e Paris-Saclay, CentraleSup\'elec, Laboratoire de G\'enie des Proc\'ed\'es et Mat\'eriaux, Gif-sur-Yvette, France \\
    $^5$Columbia Astrophysics Laboratory,~Columbia University,~New York, NY10027, U.S.A.   \looseness=-3 % avoid linesbreaks in affiliations
}

\title{ \ce{Dissociative electron attachment to the HNC$_3$ molecule}
}

\date{\today}% It is always \today, today,

\begin{abstract}
Dissociative electron attachment (DEA) to the HNC$_3$ is modeled theoretically using a first-principles approach. In HNC$_3$+$e^-$ collisions, there is a low-energy resonance, which has a repulsive character along the H+NC$_3$ coordinate and becomes a bound electronic state of the HNC$_3^-$ anion near the equilibrium of HNC$_3$. The anion state dissociates without a potential barrier towards C$_3$N$^-$+H. The cross section and the rate coefficient of the process were computed. The obtained rate coefficient at low temperatures is $5\times 10^{-9}$cm$^3$/s at 300~K. Such a value of the DEA rate coefficient makes the DEA process by three orders of magnitude more efficient in producing negative molecular ions in the interstellar space than the radiative electron attachment (REA). It is suggested that negative molecular carbon-chain ions, observed in the interstellar medium, are produced by DEA rather than REA.
%The present results combined can explain the observed abundance of C$_3$N$^-$ in the interstellar and circumstellar enviroments as due to DEA to HNC$_3$.
\end{abstract}

\maketitle

%\tableofcontents

\section{Introduction}

Dissociative electron attachment to polyatomic ions plays an important role in molecular plasma. Knowing rate coefficients and products of the reaction are important for fundamental science, such as for understanding processes in the interstellar medium and star formation, planetary atmospheres, and other applications, such as modeling the behavior of plasma in technological processes. Experiments can provide some of the required data, but they are expensive and often are not feasible, for example, if the molecule of interest is in a radical or an excited state, which is unstable in collisions with other species present nearby. Theory is now able to predict the rate coefficients with quite good accuracy and the product distribution of the DEA reaction for diatomic molecules. For polyatomic molecules, several approaches have been developed and applied to a number of molecules; however, these approaches are often limited to one particular molecule and are difficult to adapt to other molecules. In this study, we developed a quite general method to determine the DEA rate coefficients. The method is based on first principles only and can be applied to a wide range of polyatomic molecules.

Several negative molecular ions have been observed in interstellar and circumstellar clouds,  C$_n$N$^-$($n=1,3,5$), C$_m$H$^-$($m=4,6,8$) and, possibly,  \ch{C10H-} or \ch{C9N-},  \cite{C3Nm-Thaddeus-2008,C4Hm-Cernicharo2007,C5Nm-Cernicharo2008,C6Hm-McCarthy2006,C8Hm-Bruenken2007,cernicharo2023discovery,remijan2023astronomical,pardo2023new,agundez2023abundance}. Following an earlier theoretical prediction \cite{herbst81,herbst82,millar07,herbst08} that such ions can be formed by the process of radiative electron attachment (REA) and indeed could be observed in the interstellar medium (ISM), it was generally accepted by the astronomical community since the first observations of the anions in the ISM that they are formed by REA. However, accurate first-principles calculations \cite{douguet13,douguet15,khamesian16,khamesian2016b,Miguel-REA-PRA-2019.99.033412,Miguel-Carbon-Chain-ESC-2019,forer2023radiative} from two independent groups have demonstrated that the REA process is so slow that the observed relative abundances of the anions and the corresponding neutral molecules exclude the REA as the main path of anion formation. This led to the idea, suggested in several studies, that molecular anions could instead be formed in the ISM via DEA, a process generally much faster than REA at low collision energies, if it is allowed energetically. For example, \citet{petrie97} have suggested that  C$_3$N$^-$ could be formed in the ISM by DEA to the HNC$_3$ molecule. DEA is also considered as the main mechanism of molecular anion formation in Titan's atmosphere \cite{vuitton2009negative,dobrijevic20161d}. Despite this and later suggestions that DEA could be the mechanism of the molecular anion formation in the ISM, there have been no significant attempts to calculate or measure DEA rate coefficients for molecules that could produce the anions observed in the ISM. Our study addresses this gap by providing a first-principles modeling of DEA to HNC$_3$, observed in the ISM \cite{kawaguchi1992detection,vastel2018high,cernicharo2020discovery}, which forms the C$_3$N$^-$ ion at low temperatures.

This article is organized as follows. Section \ref{sec:el_structure} gives the electronic structure of HNC$_3$ and HNC$_3^-$, and  Section \ref{sec:XS} sketches the employed theoretical approach to compute the DEA cross section and rate coefficient. Section \ref{sec:discussion} discusses the obtained results. Section \ref{sec:conclusion} concludes this study.

\section{Electronic structure of HNC$_3$ and HNC$_3^-$}
\label{sec:el_structure}

\begin{figure}
	\includegraphics[scale=0.09]{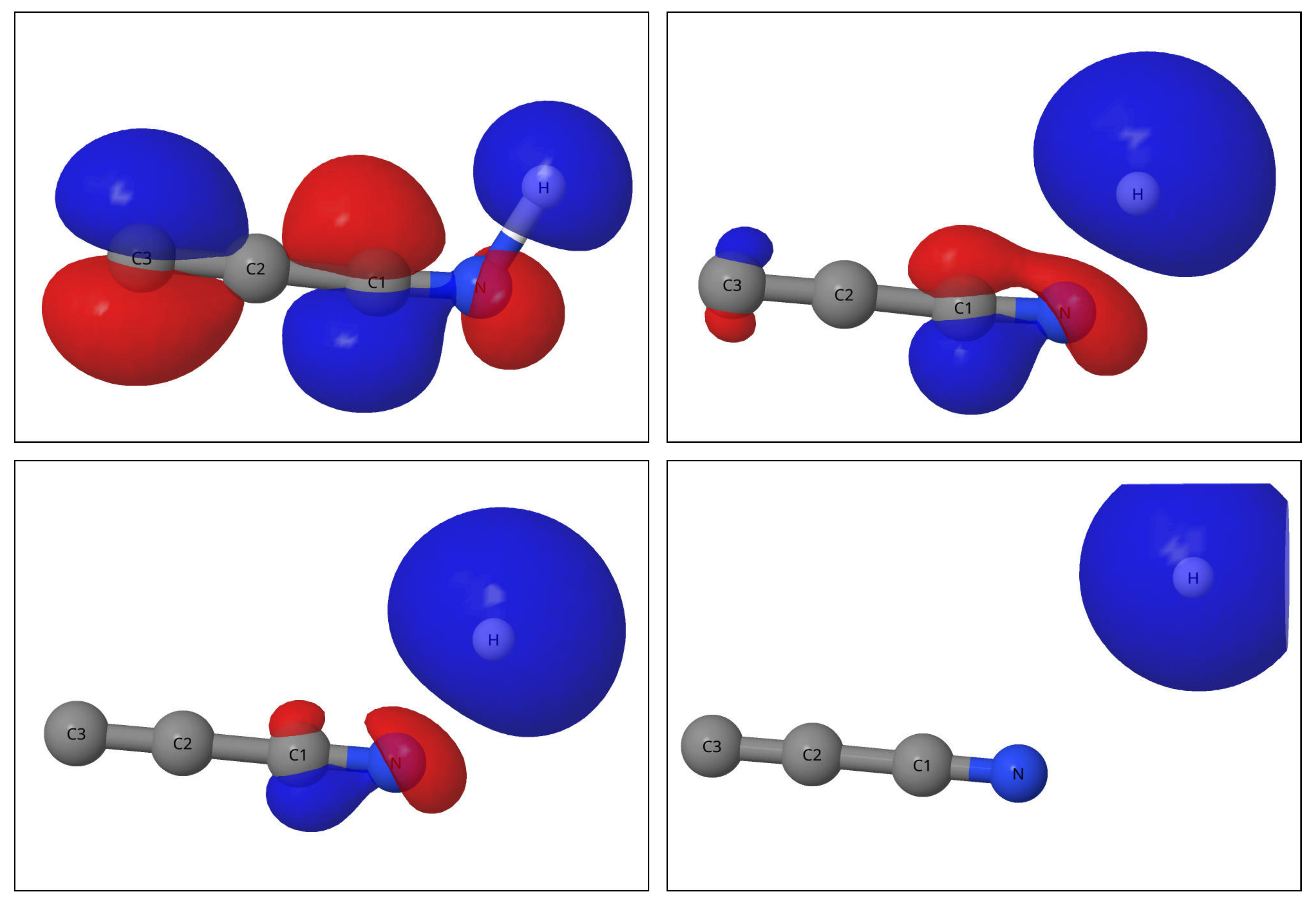}
	\caption{The evolution of the highest (singly) occupied (HOMO) natural orbital of the HNC$_3^-$ system for four geometries, where all internuclear coordinates were fixed at the values of the HNC$_3$ equilibrium, except the distance ($r_\mathrm{HN}$) between H and N: on the top left, $r_\mathrm{HN}=2.5$ $a_0$; top right, $r_\mathrm{HN}=3$ $a_0$; bottom left, $r_\mathrm{HN}=3.5$ $a_0$; bottom right, $r_\mathrm{HN}=6$ $a_0$.
	}
	\label{fig:orbitals}
\end{figure}

The HNC$_3$ molecule in its ground electronic state is planar. Its electronic state is of the $A'$ irreducible representation (irrep) of the C$_\text{s}$ point group with the electronic configuration of 11  $a'$ orbitals  and 2 $a''$ orbitals. Its equilibrium geometry has been determined by \citet{botschwina2003spectroscopic}, shown in the top left panel of Fig.~\ref{fig:orbitals}. In this study, we have also performed the electronic structure calculation using a slightly larger basis set than that of ~\citet{botschwina2003spectroscopic} and two different methods: a coupled-cluster method CCSD(T), and the multireference configuration interaction (MRCI) method, both employing the \texttt{MOLPRO} suite \cite{Molpro2012}. We have obtained results that are essentially identical to those of Ref.~\cite{botschwina2003spectroscopic}. The normal mode harmonic energies of the molecule have also been computed in this study and are given in Table~\ref{tab:modes}. Animated representations of the modes are provided in the supplementary material.

\begin{table}[h!]
\centering
\begin{tabular}{|cc|c|c|}
\hline
\multicolumn{2}{|c|}{\textbf{irrep}} & \textbf{type} & \textbf{frequency} \\
\hline
1,& $A'$ &in-plane C-C-C bending&  152 \\
2,& $A''$ &off-plane C-C-C bending &  162  \\
3,& $A'$ &C-N-H bending + N-H stretching & 470  \\
4,& $A''$ &off-plane C-C-N bending& 555 \\
5,& $A'$ &in-plane C-C-N bending& 574 \\
6,& $A'$ & C-C-C-N symmetric stretching& 943 \\
7,& $A'$ &C2-C3 and C3-N stretching& 1910  \\
8,& $A'$ & C-C-C-N antisymmetric stretching&  2251  \\
9,& $A'$ &N-H stretching&  3734 \\
\hline
\end{tabular}
\caption{\label{tab:modes}HNC$_3$ normal modes and their frequencies (in cm$^{-1}$), obtained using the CCSD(T) method with the cc-pVQZ basis set.}
\end{table}

\begin{figure}
	\includegraphics[scale=0.25]{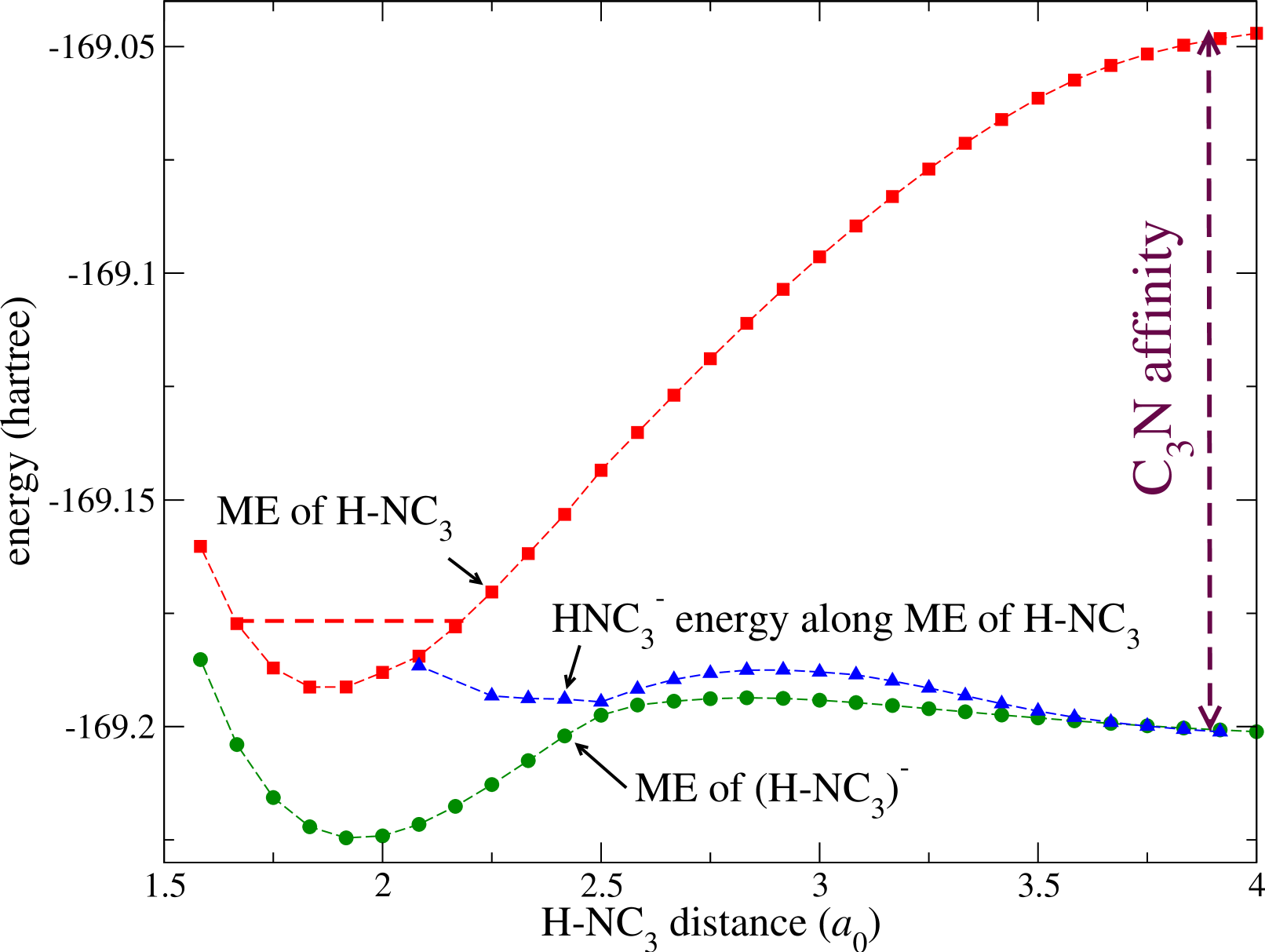}
	\caption{Energy diagram of the neutral molecule HNC$_3$ and the anion HNC$_3^-$ as a function  $r_\mathrm{HN}$. The upper (red squares) and lower (green circles) curves show the minimum energy (ME) of HNC$_3$ and HNC$_3^-$, respectively, obtained for a given value of $r_\mathrm{HN}$ while all other internuclear coordinates are varied. Notice that although the $r_\mathrm{HN}$ distance is the same for each value of the $x$ coordinate for the two sets (circles and squares), other internuclear distances are not the same for the two data sets. The blue triangles show the energy (not the minimum energy) of the HNC$_3^-$ computed for the geometry obtained for HNC$_3$.
	}
	\label{fig:MEP}
\end{figure}

Considering the  HNC$_3$+$e^-$ system at the geometry of the HNC$_3$ equilibrium, there is an electronic resonance of the $A'$ and of the $A''$ irreps at 0.142~eV and 1.064~eV (scattering energy), respectively. The lowest dissociation limit of the $A'$ resonant state corresponds to C$_3$N$^-$+H. Figure \ref{fig:MEP} demonstrates energies of the HNC$_3$ molecule and the HNC$_3^-$ anion as a function of the distance $r_\mathrm{HN}$ between the H and N atoms.  As evident from the figure, the geometry where the  HNC$_3^-$ PES approaches and crosses the PES of HNC$_3$ is near $r_\mathrm{HN}=2.06$ $a_0$ (here and below $a_0$ is the Bohr radius, the atomic unit of length). Figure  \ref{fig:orbitals} shows how the highest singly occupied orbital of  HNC$_3^-$ changes when the distance $r_\mathrm{HN}$ increases. It changes from an $a'$ orbital that is delocalized along the molecule near the HNC$_3$ equilibrium to the 1$s$ orbital that is localized on the hydrogen atom at large values of $r_\mathrm{HN}$.

\begin{figure}
	\includegraphics[scale=0.28]{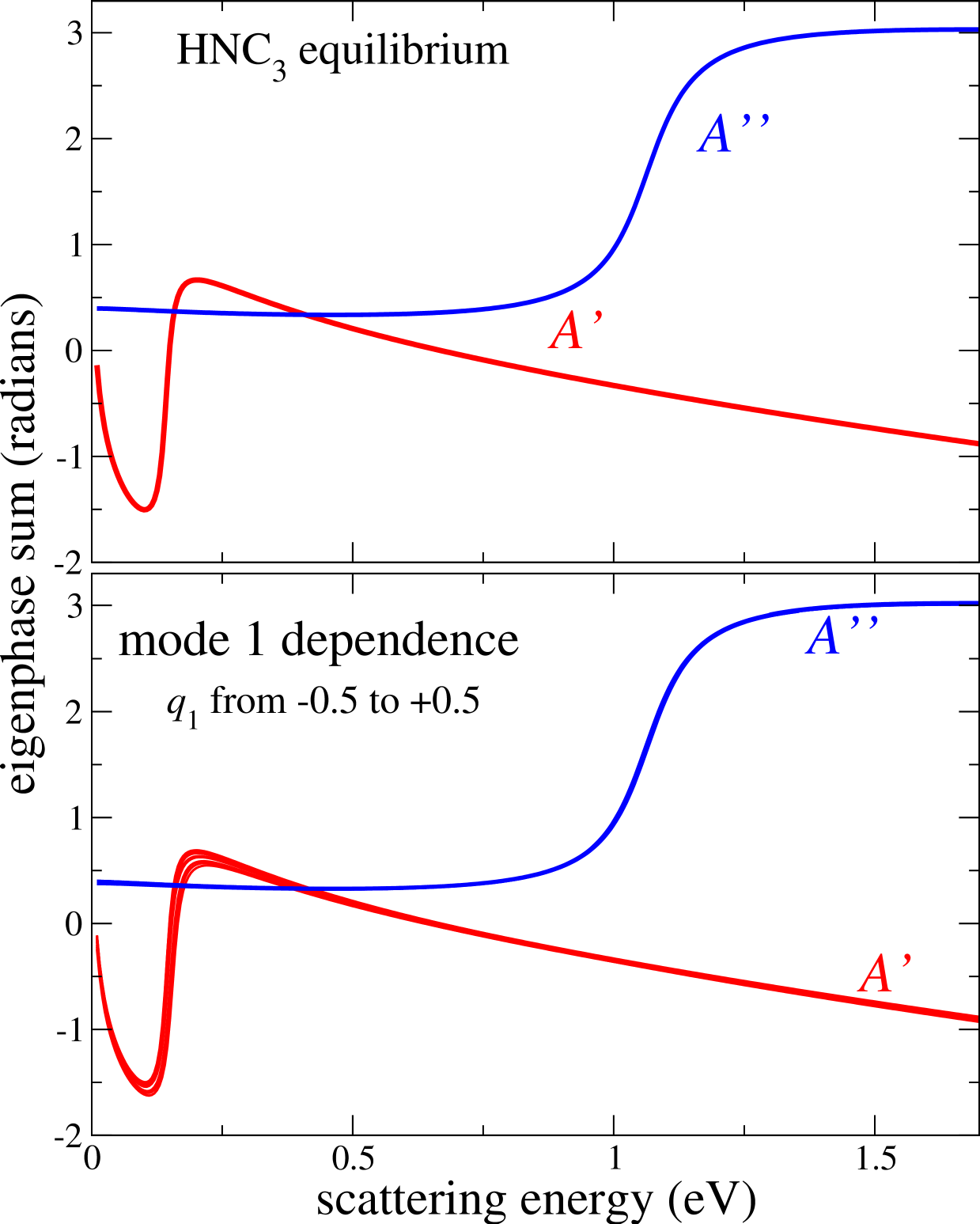}
	\caption{The upper panel shows $e^- +$ HNC$_3$ scattering eigenphase sums  for the two symmetries $^2A'$ and $^2A''$ obtained for the geometry of HNC$_3$ equilibrium. The lower panel shows the eigenphase sums for several displacements $q_1$ along mode 1, in the interval of $q_1$ from -0.5 to 0.5. As one can see, varying $q_1$ in this interval does not change significantly positions and widths of the resonances.}
	\label{fig:mode1}
\end{figure}

To model the DEA process, we have modified the approach developed by \citet{Yuen2019}, which is based on the DEA theory for diatomic molecules \cite{omalley66}. The approach requires the knowledge of the HNC$_3$ PES near its equilibrium position, the HNC$_3$ vibrational modes, as well as the PES and autodetachment widths of the anion near the HNC$_3$ equilibrium and along the path to the dissociation of HNC$_3^-$. The  HNC$_3$ PES and its vibrational modes were calculated as discussed above. Similarly to Ref.~\cite{Yuen2019}, for determination of the anion PES and widths, the UK R-matrix \cite{carr12} and the Quantemol \cite{tennyson07quantemol} codes were employed to perform $e^-$-HNC$_3$ calculations for several geometries of the target molecule near the HNC$_3$ equilibrium and towards the dissociation limit of HNC$_3^-$. The $A'$ and $A''$ resonances, mentioned above, are visible in Fig.~\ref{fig:mode1} (upper panel), showing the $A'$ and $A''$ eigenphase sums. The $A'$ resonance becomes a bound electronic state of HNC$_3^-$ for small displacements, typically, involving a stretching of the H-N bond. Figure \ref{fig:MEP} shows the energy of the bound state as a function of the H-N bond length, while all other internuclear distances are configured such that the HNC$_3$ energy is minimized. The $A'$ resonance plays a crucial role in the electron attachment at low collision energies and following dissociation, while the higher-energy $A''$ resonance is not active in the process at temperatures below 1000~K.
%Below, if not mentioned otherwise, we discuss only the $A'$ resonance.

\begin{figure}
	\includegraphics[scale=0.28]{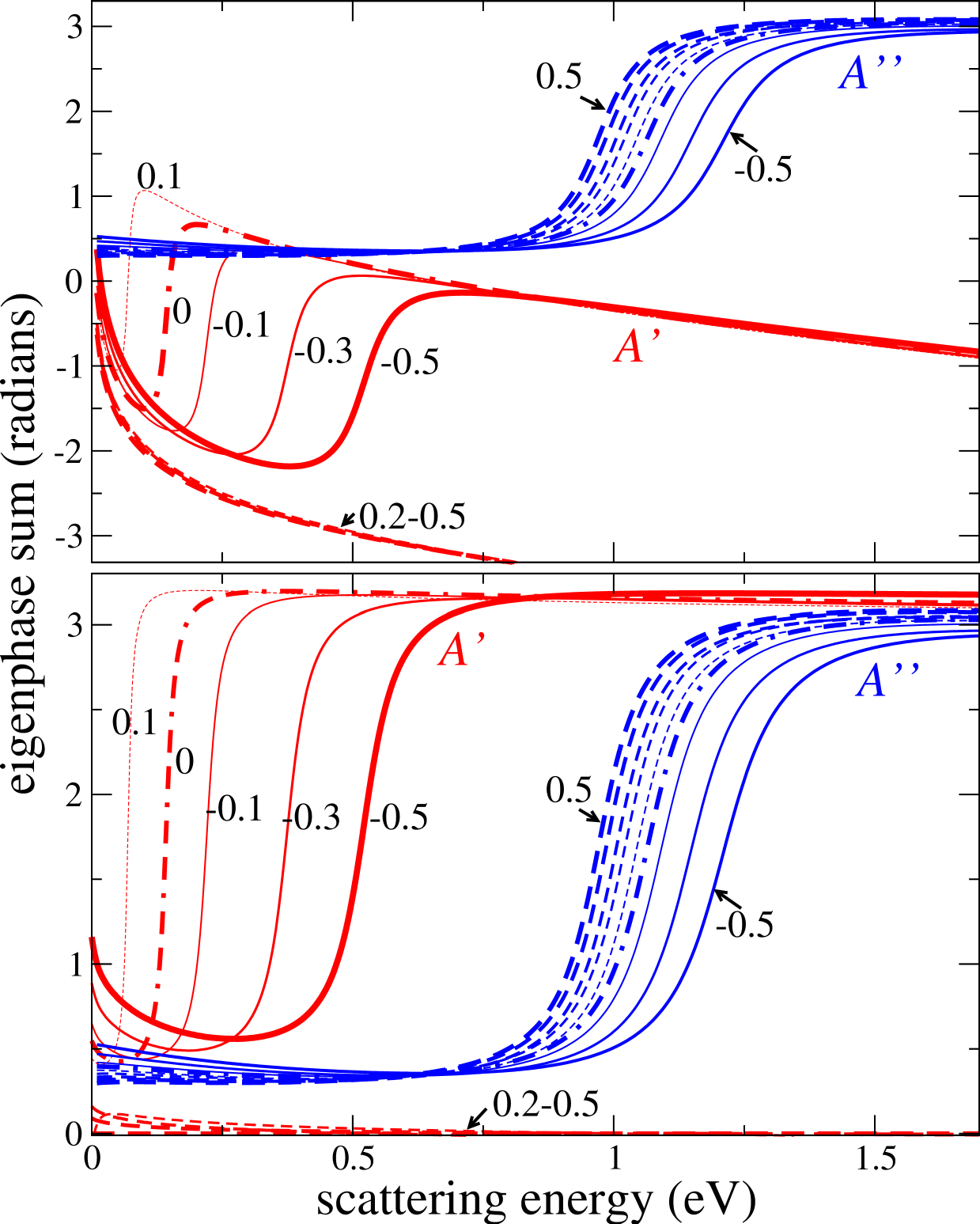}
	\caption{Eigenphase sums for mode 3. The upper panel shows the raw eigenphase sums. The lower panel shows the same data with the non-resonant contribution subtracted for the $^2A'$ symmetry. The subtracted $^2A'$ non-resonant contribution was evaluated at displacement $q_3=0.5$.
	}
	\label{fig:mode3}
\end{figure}

The lower panel of Fig.~\ref{fig:mode1} shows the eigenphase sums for different displacements along mode 1 near the equilibrium geometry of HNC$_3$. Similarly to Ref.~\cite{Yuen2019}, dimensionless normal mode coordinates are used, i.e. coordinates in units of the harmonic oscillator length $\sqrt{\hbar/m\omega}$, where $m$ is the mass and $\omega$ is the frequency of the oscillator. As evident from the figure, energies and widths of the two resonances do not change significantly in the interval $q_1\in [-0.5;0.5]$ -- compared, for example, to mode 3, which is demonstrated in the upper panel of Fig.~\ref{fig:mode3}. At displacement $q_3\sim0.15$, the $^2A'$ resonance becomes a bound HNC$_3^-$ state. Performing the calculations for all 9 normal modes, it was found that for modes 2 and 4 the energy and width of the $^2A'$ resonance also do not change. In contrast, for all other modes 5-9, the dependence is significant and should be accounted for in the DEA process. In the following discussion, modes 1, 2, and 4 will be referred to as inactive modes, and modes 3, 5-9 as active modes.

There are resonant and potential scattering (non-resonant) contributions to the scattering phases and the eigenphase sums. The resonant contribution changes significantly with $q$ for active modes, while the potential scattering one does not. One can separate the resonant contribution, subtracting the eigenphase sum for a geometry where the resonant contribution is not present, i.e. where the resonance becomes a bound state. The lower panel of Fig.~\ref{fig:mode3} shows the same data as the upper panel with the potential contribution subtracted for $^2A'$ symmetry. The subtracted potential contribution is the total eigenphase sum, evaluated at displacement $q_3=0.5$.

\begin{figure}
	\includegraphics[scale=0.28]{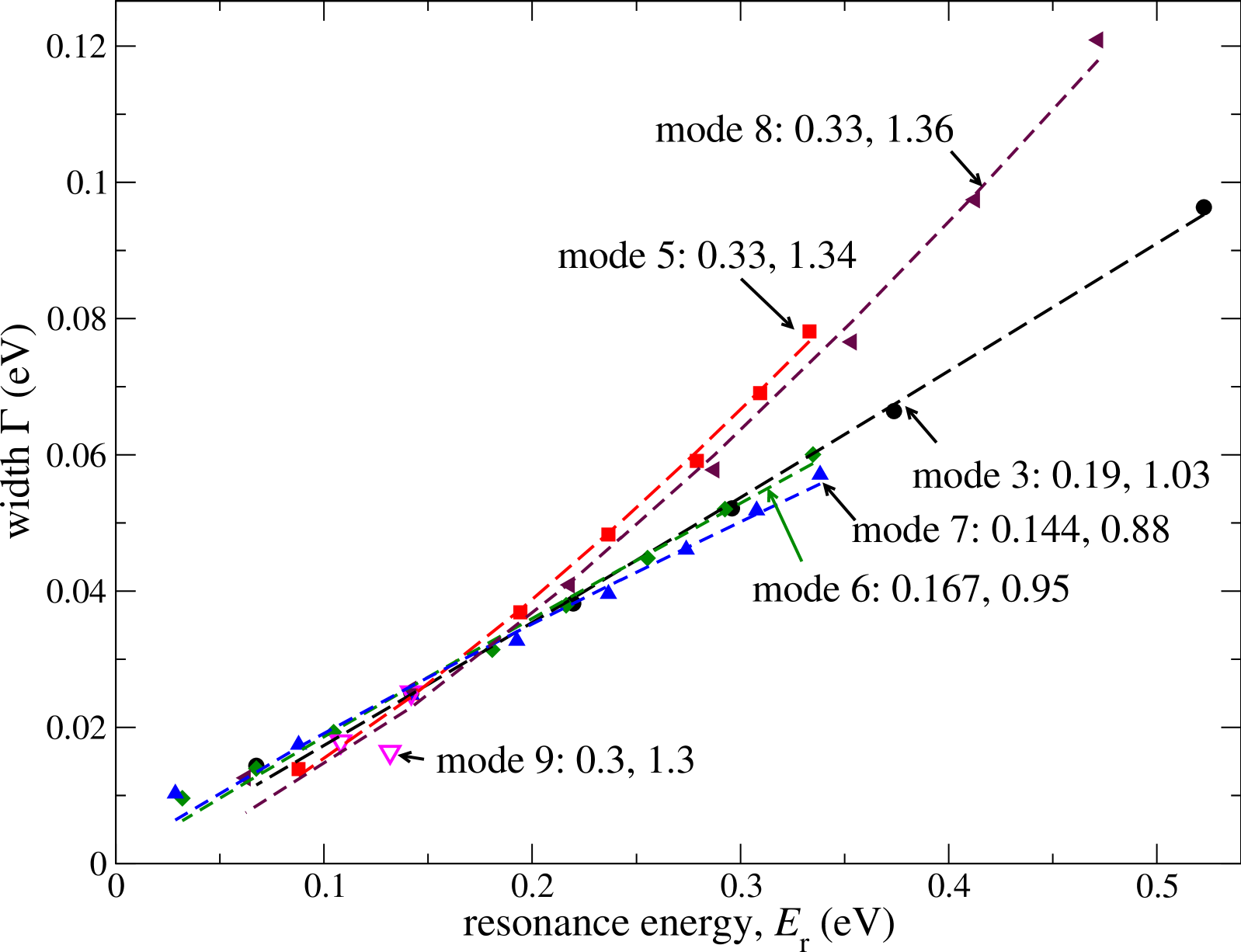}
	\caption{Dependence of the $^2A'$ resonance width $\Gamma(q_i)$ on its energy $E_\mathrm{r}(q_i)$ while the value of the normal mode coordinate $q_i$ is changing. The dependence is shown for six active modes $i=3,5-9$. The calculated values  $\Gamma(q_i)$ and $E_\mathrm{r}(q_i)$ are shown with symbols. The dependence $\Gamma(E_\mathrm{r})$ for each mode is fitted with the power law of
	Eq.~(\ref{eq:gamma_vs_E}) (dashed lines). 	The fitted parameters $a$ and $b$ are indicated on the figure. }
	\label{fig:gamma_vs_energy_all_modes}
\end{figure}

Figure~\ref{fig:gamma_vs_energy_all_modes} summarizes energies and widths of the $A'$ resonance as a function of the displacements for the active modes. Symbols mark the values obtained from the eigenphase sums, while the lines are analytical fits to the power law \cite{Yuen2019}
\begin{equation}
\label{eq:gamma_vs_E}
 \Gamma(E_\mathrm{r})=aE_\mathrm{r}^b\,.
\end{equation}
Together with the PES of the neutral molecule, represented by the quadratic potential along the normal modes, the data shown in figure represent the PES surface $U(\vec q)$ and its autodetachment width in the region near HNC$_3$ minimum and where the PES's of  HNC$_3$ and HNC$_3^-$ cross each other.

\section{Cross section and rate coefficient}
\label{sec:XS}

 Similarly to Ref.~\cite{Yuen2019}, to account for the multidimensional nature of the nuclear motion, one determines the dissociative coordinate $\vec s$ pointing in the direction of the steepest descent along the surface $U(\vec q)$: $ \vec s =-{\vec \nabla U}/{|\vec \nabla U|}$. In the present DEA model, the scattering electron, incident on \target, is captured in the $^2A$ resonant state. Once it is captured, the nuclear motion proceeds along the $U(\vec q)$ PES in the direction of $\vec s$.

 For a displacement $ds$ along  $\vec s$, normal coordinate displacements $dq_i$ change as
\begin{equation}
dq_i=-\frac{1}{|\vec \nabla U|}\frac{\partial U}{\partial q_i}ds\,,
\end{equation}
allowing us to compute the displacements $dq_i$, the normal and Cartesian coordinates of the geometries for motion on $U(\vec q)$ along the direction $\vec s$.

\begin{figure}
	\includegraphics[scale=0.28]{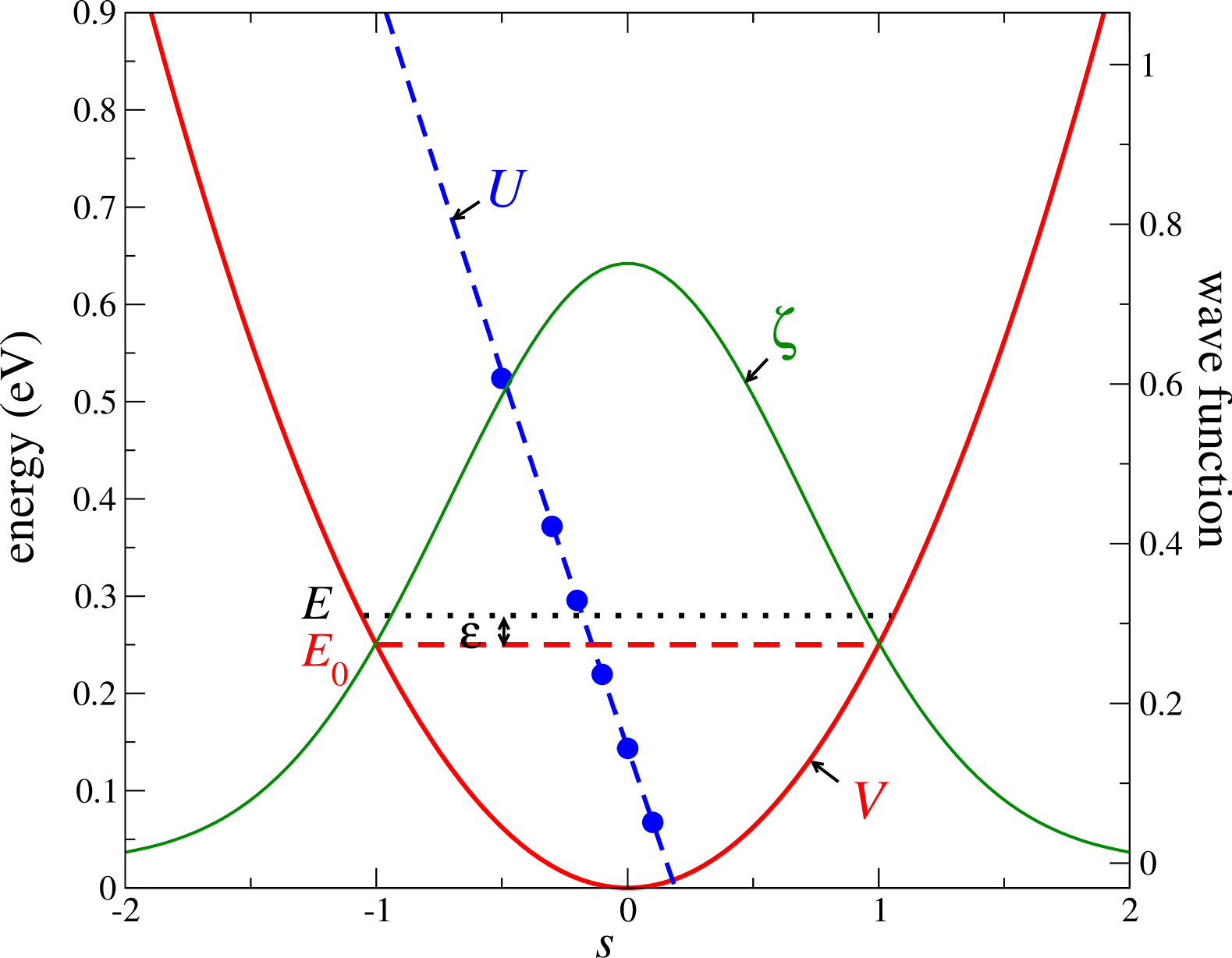}
	\caption{Schematic view of the DEA process along the coordinate of the steepest descent. The actual process takes place in the 9-dimensional space of the internuclear coordinates.
	}
	\label{fig:energy_vs_s}
\end{figure}

The process of DEA is schematically shown in Fig.~\ref{fig:energy_vs_s}, representing the one-dimensional cut along the coordinate of the steepest descent $s$. In the figure, $E$ is the total energy of the $e^-+$HNC$_3$ system. The initial vibrational state $\zeta(s)$ of \target~ with energy $E_0$ is shown with the green line, while the dissociating state $U(s)$ of \anion~ is shown by blue circles (calculated) and the dashed line (analytical fit).
The electron, incident on \target~ with asymptotic energy $\varepsilon=E-E_0$, is captured in the $^2A'$ resonant state. After the electron capture, the nuclear motion proceeds along the $U(\vec q)$ PES in the direction of $\vec s$, as shown in Fig.~\ref{fig:gamma_vs_energy_all_modes}. The DEA cross section is computed similarly to Refs.~\cite{omalley66,Yuen2019} as
\begin{equation}
\sigma (\varepsilon) = \frac{2\pi^2}{k^2}  \frac{ \Gamma(s_{\varepsilon})}{|U'(s_E)|}  |\zeta (s_E)|^2,
\label{eq:xsec}
\end{equation}
where the classical turning point $s_E$ for $U(s)$ and the Frank-Condon point $s_{\varepsilon}$
are obtained by solving $U(s_E) = E$ and $E_\mathrm{r}(s_{\varepsilon}) = \varepsilon$, respectively. Notice that $U(\vec q)=V(\vec q)+E_\mathrm{r}(\vec q)$ for any geometry $\vec q$.

\begin{figure}
	\includegraphics[scale=0.4]{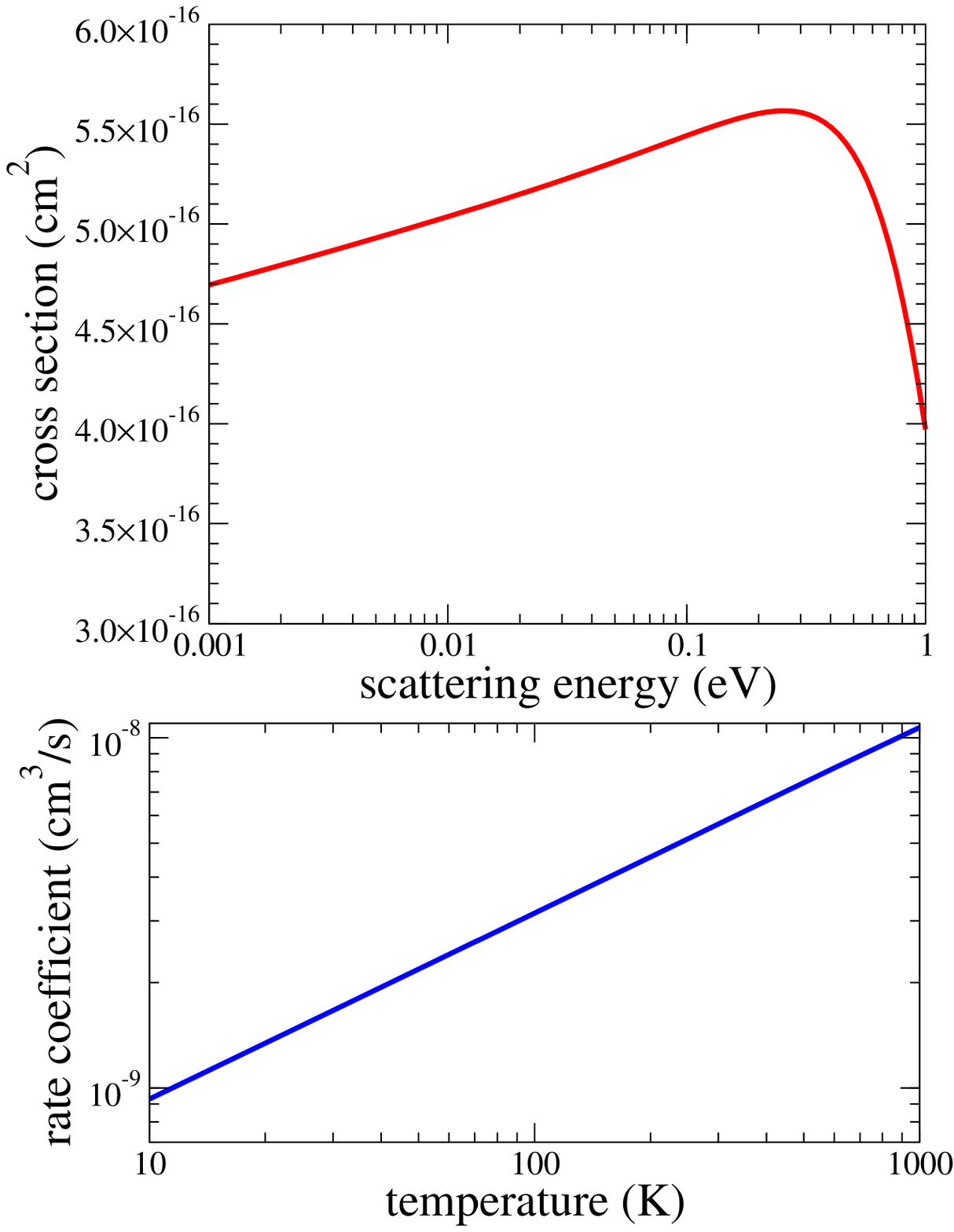}
	\caption{Cross section and rate coefficient for DEA to HNC$_3$}
	\label{fig:CS}
\end{figure}

The resulting DEA cross section, obtained using Eq.~\ref{eq:xsec}, and the corresponding thermally averaged rate coefficient are plotted in Fig.~\ref{fig:CS}. The cross section does not change significantly, varying between 4 and 5.6 \AA$^2$ in the interval of scattering energies from 1~meV to 1~eV. As a result, the rate coefficient is almost an exact square root function  of temperature: $\tilde k\sqrt{T/\textrm{K}}$ with $\tilde k=3.2(\pm 0.1)\times 10^{-10}$cm$^3$/s.

\section{Discussion}
\label{sec:discussion}
Equation (\ref{eq:xsec}) for the DEA cross section does not include the autodetachment factor, which decreases the cross section. However, in the present case the autodetachment probability is negligible because (1) the time $\tau_d$, needed for the system to slide along the coordinate $s$ to the geometries where autodetachment is forbidden, is much smaller than the autodetachment time $\tau_a=\hbar/\Gamma(s_{\varepsilon})$ and (2) the dissociation is energetically open for any scattering energy (see Fig.~\ref{fig:MEP}). Time $\tau_d$ can be estimated from the the distance $\Delta s$ between the Franck-Condon point  $s_{\varepsilon}$ and the point where the PESs of the ion and the neutral molecule cross. Inspecting Fig.~\ref{fig:energy_vs_s}, one estimates $\Delta s$ to be about 0.1 of the oscillator length, i.e. the time needed for the nuclei to move this distance is of the order of 0.1 of the period of oscillations (or smaller), $\tau_d \lesssim 0.2\pi/\omega$. Taking an average harmonic frequency $\omega_5$, we obtain the ratio
\begin{equation}
 \tau_d/\tau_a\lesssim 0.2\pi\Gamma(s_{\varepsilon})/\hbar\omega_5%\approx 0.6*0.005/0.07=0.04
\end{equation}
For scattering energy 25~meV, corresponding to the room temperature, the ratio is about 0.04, i.e. the autodetachment probability is about 0.04. It is below the expected uncertainty of the present model and, therefore, can be ignored.

The obtained DEA cross section and the rate coefficient are  significantly larger, by about 4 orders of magnitude, than the values obtained for H$_2$CN \cite{Yuen2019}. It is because in the DEA  of H$_2$CN, the PESs of the ion and the neutral molecule cross far from the equilibrium geometry of H$_2$CN and widths of the H$_2$CN$^-$ resonant state are by about an order of magnitude smaller than the HNC$_3^-$ widths.

In the present theoretical approach, the rotational motion of the target molecule was neglected. Although accurate calculations have not been performed yet, it is expected that the rotational excitation of HNC$_3$ during the process should increase and, probably, significantly the DEA cross section. The target molecule has a relatively large dipole moment of 2.4~$ea_0$. Therefore, the cross section for rotational excitation by electron impact is expected to be significant at scattering energies above the rotational excitation energy. If the scattering energy is below the energy threshold for rotational excitation of the target, the scattering electron will not autoionize immediately and will orbit the molecule, approaching it another or several more times, thereby increasing the DEA cross section by the factor corresponding to the number of approaches during such a rotational resonance.  Our preliminary estimates show that the factor is significant, but more accurate calculations are needed. A dedicated study of the role of rotational resonances in the DEA of molecules, leading to the formation of the anions observed in the ISM, will follow.

The  DEA cross section and the rate coefficient, obtained above without accounting for the rotational motion, should be considered as a lower-bound estimate. Nevertheless, it is useful to compare the efficiency of formation of C$_3$N$^-$ by DEA to \target~with radiative electron attachment (REA) to C$_3$N using the obtained DEA rate coefficient.
Taking the value of 500 for the ratio of C$_3$N to \target~abundances derived from the IRC+10216 source \cite{C3Nm-Thaddeus-2008} and the rate coefficients for formation of C$_3$N$^-$ by REA and DEA, $k_\mathrm{REA}=5\times 10^{-15}$cm$^3$/s \cite{khamesian16} and  $k_\mathrm{DEA}=5\times 10^{-9}$cm$^3$/s (Fig.~\ref{fig:CS}), respectively, at 300~K, we obtain the following ratio of formation efficiencies:
\begin{equation}
  \frac{[e^-][\mathrm{HNC_3}]k_\mathrm{DEA}}{[e^-][\mathrm{C_3N}]k_\mathrm{REA}} \approx 2000\,.
\end{equation}
Thus, the DEA mechanism is estimated to be about 2000 times more efficient in producing C$_3$N$^-$ than REA at 300~K in IRC+10216.

\section{Conclusion}
\label{sec:conclusion}
In this study, we have considered dissociative electron attachment to the HNC$_3$. The approach combines the electron-scattering calculations and the O'Malley theory of dissociative attachment generalized to polyatomic targets. In HNC$_3$+$e^-$ collisions, there is a low-energy resonance, which has a repulsive character along the H+NC$_3$ coordinate and becomes a bound electronic state of the HNC$_3^-$ anion near the equilibrium of HNC$_3$. The anion state dissociates without a potential barrier. The present {\i ab initio} calculations, performed without accounting for the rotational motion of HNC$_3$ have demonstrated that the DEA is by several orders of magnitude more efficient in formation of C$_3$N$^-$ anion in the ISM than the REA, previously considered by the astronomical community as the main mechanics of anion formation in the ISM. Other molecular ions observed in the ISM could also be formed by DEA to the corresponding neutral molecules.

\section*{Acknowledgements}
%This work acknowledges support from the National Science Foundation, Grant Nos.

VK and JF acknowledge the support from the US National Science Foundation, grants 2409570 and 2303895 respectively. The study was also partially supported by the Transatlantic Mobility Program and Chateaubriand Fellowship of the Office for Science and Technology of the Embassy
of France in the United States, Programme National “Physique et Chimie du Milieu Interstellaire” (PCMI) of CNRS/INSU, the program “Accueil des chercheurs étrangers” of CentraleSupélec and from French State aid under France
2030 (QuanTEdu-France) bearing the reference ANR-22-CMAS-0001 and PEPR - SPLEEN Plasma-N-Act.

\bibliography{DEA-REA}

\pagebreak

\newpage

% \begin{figure}
% 	\includegraphics[scale=0.25]{figs/12a'_3rhn.jpg}
% 	\caption{12a'-3rhn
% 	}
% 	\label{fig:orbital_12a'_3rhn}
% \end{figure}
%
%
% \begin{figure}
% 	\includegraphics[scale=0.25]{figs/12a'_3.5rhn.jpg}
% 	\caption{12a'-3.5rhn
% 	}
% 	\label{fig:orbital_12a'_3.5rhn}
% \end{figure}
%
%
%
% \begin{figure}
% 	\includegraphics[scale=0.17]{figs/12a'_4rhn.jpg}
% 	\caption{12a'-4rhn
% 	}
% 	\label{fig:orbital_12a'_4rhn}
% \end{figure}
%
% \begin{figure}
% 	\includegraphics[scale=0.25]{figs/12a'_6rhn.jpg}
% 	\caption{12a'-6rhn
% 	}
% 	\label{fig:orbital_12a'_6rhn}
% \end{figure}

%
% \begin{figure}
% 	\includegraphics[scale=0.28]{figs/mode2.pdf}
% 	\caption{Mode 2
% 	}
% 	\label{fig:mode2}
% \end{figure}
%
% \begin{figure}
% 	\includegraphics[scale=0.28]{figs/mode4.pdf}
% 	\caption{Mode 4
% 	}
% 	\label{fig:mode4}
% \end{figure}
% \begin{figure}
% 	\includegraphics[scale=0.28]{figs/mode5.pdf}
% 	\caption{Mode 5
% 	}
% 	\label{fig:mode5}
% \end{figure}
% \begin{figure}
% 	\includegraphics[scale=0.28]{figs/mode6.pdf}
% 	\caption{Mode 6
% 	}
% 	\label{fig:mode6}
% \end{figure}
% \begin{figure}
% 	\includegraphics[scale=0.28]{figs/mode7.pdf}
% 	\caption{Mode 7
% 	}
% 	\label{fig:mode7}
% \end{figure}
% \begin{figure}
% 	\includegraphics[scale=0.28]{figs/mode8.pdf}
% 	\caption{Mode 8
% 	}
% 	\label{fig:mode8}
% \end{figure}
% \begin{figure}
% 	\includegraphics[scale=0.28]{figs/mode9.pdf}
% 	\caption{Mode 9
% 	}
% 	\label{fig:mode9}
% \end{figure}

\end{document}